\documentclass[conference]{IEEEtran}
\IEEEoverridecommandlockouts
\usepackage{amsmath,amssymb,amsthm}
\usepackage{graphicx}
\usepackage{cite}
\usepackage{tikz}
\usetikzlibrary{positioning,calc,arrows.meta}
\usepackage[hidelinks]{hyperref}
\usepackage[nohyperlinks]{acronym}
\usepackage{float}

\newif\ifblind
\blindfalse

\makeatletter
\g@addto@macro\normalsize{%
  \setlength\abovedisplayskip{3pt plus 2pt minus 1pt}%
  \setlength\belowdisplayskip{3pt plus 2pt minus 1pt}%
  \setlength\abovedisplayshortskip{1pt plus 2pt}%
  \setlength\belowdisplayshortskip{3pt plus 2pt minus 1pt}}
\makeatother

\newtheorem{theorem}{Theorem}
\theoremstyle{definition}
\newtheorem{definition}{Definition}

\newcommand{\R}{\mathbb{R}}
\newcommand{\N}{\mathcal{N}}
\newcommand{\T}{\mathcal{T}}
\newcommand{\E}{\mathbb{E}}
\newcommand{\dd}{\mathrm{d}}
\newcommand{\X}{\mathcal{X}}
\newcommand{\summ}[1]{\textstyle\sum_{\sim\{#1\}}}
\newcommand{\bx}{\mathbf{x}}
\newcommand{\bA}{\mathbf{A}}
\newcommand{\iintT}{\iint_{\T^2}}
\DeclareMathOperator{\tr}{tr}
\acrodef{AR}{autoregressive}
\acrodef{BER}{bit error rate}
\acrodef{BP}{belief propagation}
\acrodef{BPSK}{binary phase-shift keying}
\acrodef{GaBP}{Gaussian belief propagation}
\acrodef{iid}[i.i.d.]{independent and identically distributed}
\acrodef{KFL}{Kschischang, Frey and Loeliger}
\acrodef{LMMSE}{linear minimum mean-square error}
\acrodef{MIMO}{multiple-input multiple-output}
\acrodef{PDF}{probability density function}
\acrodef{SNR}{signal-to-noise ratio}
\acrodef{SPA}{sum-product algorithm}
\acrodef{FSPA}{functional SPA}
\acrodef{ISAC}{integrated sensing and communications}
\acrodef{CSI}{channel state information}
\acrodef{DoA}{direction of arrival}
\acrodef{CAPA}{continuous aperture array}
\acrodef{BL}{bandlimited}

\begin{document}

\title{Functional Factor Graphs and the\\ Functional Sum-Product Algorithm}

\ifblind
\author{\IEEEauthorblockN{Anonymous Authors}\IEEEauthorblockA{\emph{Affiliations withheld for review}}}
\else
\author{\IEEEauthorblockN{Kuranage Roche Rayan Ranasinghe and Giuseppe Thadeu Freitas de Abreu}
\IEEEauthorblockA{\emph{School of Computer Science and Engineering, Constructor University, Bremen, Germany}\\
Email: \{kranasinghe,gabreu\}@constructor.university}
\vspace{-5ex}}
\fi

\maketitle

\begin{abstract}
We propose a generalization of factor graphs and the \ac{SPA}, widely used in the estimation of data, channel state information and sensing parameters, to unknowns that are \emph{functions}, such as waveforms, time-varying channels and continuous-aperture array currents, which must otherwise be sampled first.
In order to mitigate the challenge that functions have no \acp{PDF}, the proposed scheme measures densities relative to the prior of each function, not to the \emph{flat reference} implicit in every \ac{PDF}.
Messages then become likelihood ratios, and the factor-node rule an average over the priors, still exact on trees.
For Gaussian models, the proposed \ac{FSPA} is \ac{GaBP} with kernels in place of matrices and a grid-independent convergence condition, whereas scalar \ac{GaBP} on grid samples provably slows down under refinement, as information hops from sample to sample.
Simulations in a multi-user detection setting confirms the latter, and shows that sampling first can also fail for smooth priors.
\end{abstract}
\acresetall\acused{iid}\acused{BER}\acused{BL}

\section{Introduction}\label{sec:intro}
Factor graphs, as introduced by Kschischang, Frey and Loeliger~\cite{KschischangFreyLoeliger2001}, represent a global discrete function such as a posterior, in terms of local factors, such as priors and likelihoods, enabling the computation of marginals exactly on trees and approximately otherwise via one local rule, the \ac{SPA}, or \ac{BP}. 

As such, the \ac{SPA}, with its max-product form, unifies the forward-backward and Viterbi algorithms, Kalman filtering, iterative decoding and \ac{GaBP}~\cite{AjiMcEliece2000,Forney2001,Loeliger2004,WeissFreeman2001,Malioutov2006,Shental2008}, the latter popular in low-complexity receivers~\cite{RanasingheAFDMJCDE2025,RanasingheCAPAGaBP2026}.
In all these cases, however, variables are scalars or vectors, whereas many unknowns in communications and sensing settings are \emph{functions}, such as waveforms, time-varying channel gains, delay-Doppler spreading functions~\cite{RanasingheAFDMJCDE2025} or aperture currents~\cite{WangCAPA2026,RanasingheCAPAGaBP2026}.
The usual practice is to \emph{discretize, then factor}, i.e., to sample every function at $K$ grid points and to run the \ac{SPA}, or \ac{GaBP}, on the samples.
Does the number of parallel iterations then stay bounded under grid refinement?
We prove that it cannot, for \ac{GaBP} on grid samples, as information hops from sample to sample, and its convergence guarantee weakens or vanishes, as adjacent samples are strongly coupled.

Message passing over functions has so far been confined to trees, as in infinite-dimensional Kalman filtering~\cite{Falb1967,Curtain1975,Sarkka2013}, waveform interpolation~\cite{Bolliger2013} and continuous-aperture receivers~\cite{WangCAPA2026}, or to categorical Bayes filtering~\cite{Fritz2025}, \ac{BP} over Gaussian random fields stored at finitely many points~\cite{Seeger2007,Barfoot2014}, continuous-time~\cite{ElHay2010} and set-type~\cite{Kim2024} \ac{BP}, and dynamic cavity methods without a loopy convergence theory~\cite{Tarabolo2025,Metz2026}.
Kernel and continuous-state \ac{BP}~\cite{SongKBP2011,Noorshams2013} pass function-valued messages about finite-dimensional variables, and~\cite{SongKBP2011,terHorst2026} list \ac{GaBP} and categorical \ac{BP} in infinite dimensions as open.
Functional graphical models~\cite{QiaoGuoJames2019,WaghmarePanaretos2025}, interacting diffusions~\cite{Lacker2021} and Bayesian inverse problems, whose samplers take the prior as reference and do not degrade under refinement~\cite{Stuart2010,Cotter2013}, pass no messages, and block \ac{GaBP}~\cite{MalioutovThesis2008,Kamper2019,Du2018,Fanaskov2022} and its multigrid acceleration~\cite{ElKurdi2015,Key2025} are finite-dimensional.
An \ac{SPA} over functions, with well-defined messages and a grid-independent convergence theory for loopy \ac{GaBP}, thus remains open.

In view of the above, we propose to \emph{factor, then discretize}, i.e., to build the factor graph with one node per function, and to discretize only to compute messages.
The contributions of the article are \emph{functional factor graphs}, the \ac{FSPA} and their three implications (Section~\ref{sec:ffg}), and \emph{functional \ac{GaBP}}, whose convergence condition is grid-independent, unlike \ac{GaBP} on grid samples, which provably slows down under refinement (Section~\ref{sec:gauss}), with a sparse multi-user example in which grid \ac{GaBP} with the exact prior diverges and a direct solve is costly (Section~\ref{sec:num}).

\noindent\emph{Notation:} Boldface denotes finite-dimensional vectors and matrices; functions and kernels carry their arguments, e.g., $x(t)$ and $A(t,t')$, with $t,t'$ in a compact domain $\T$; $L^2(\T)$ is the set of finite-energy functions on $\T$; $\delta(\cdot)$ is the Dirac delta; $\rho(\cdot)$ is the spectral radius; primes denote further objects of the same kind; and arguments are dropped when clear, e.g., $\iintT zWz\triangleq\iintT z(t)W(t,t')z(t')\,\dd t\,\dd t'$.

\section{Factor Graphs and the \texorpdfstring{\acs{SPA}}{SPA}}\label{sec:kfl}
Following~\cite{KschischangFreyLoeliger2001}, consider a global function of the variables $x_1,\dots,x_I$, typically an unnormalized posterior, i.e., a product of priors and likelihoods, which factors as
\begin{equation}\label{eq:g}
g(x_1,\dots,x_I)=\textstyle\prod_{m=1}^{M}f_m(X_m),
\end{equation}
where $X_m$ is the set of arguments of the local function $f_m$.

The factor graph of the global function $g$ has a factor node for each $f_m$ and a variable node for each $x_i$, joined by an edge if and only if $x_i\in X_m$.
The \ac{SPA} computes the marginals of $g$ by passing messages along the edges with the rules
\begin{align}
\mu_{x\to f}(x)&=\textstyle\prod_{f'\in n(x)\setminus\{f\}}\mu_{f'\to x}(x), \label{eq:v2f}\\
\mu_{f\to x}(x)&=\summ{x}\Big(f(X)\textstyle\prod_{x'\in n(f)\setminus\{x\}}\mu_{x'\to f}(x')\Big), \label{eq:f2v}
\end{align}
where $f\triangleq f_m$ has the arguments $X\triangleq X_m\ni x$; $\mu_{x\to f}$ and $\mu_{f\to x}$ are the messages from $x$ to $f$ and back, both functions of $x$; $n(\cdot)$ denotes the neighbors of a node, with $f'$ and $x'$ the other neighbors of $x$ and $f$.

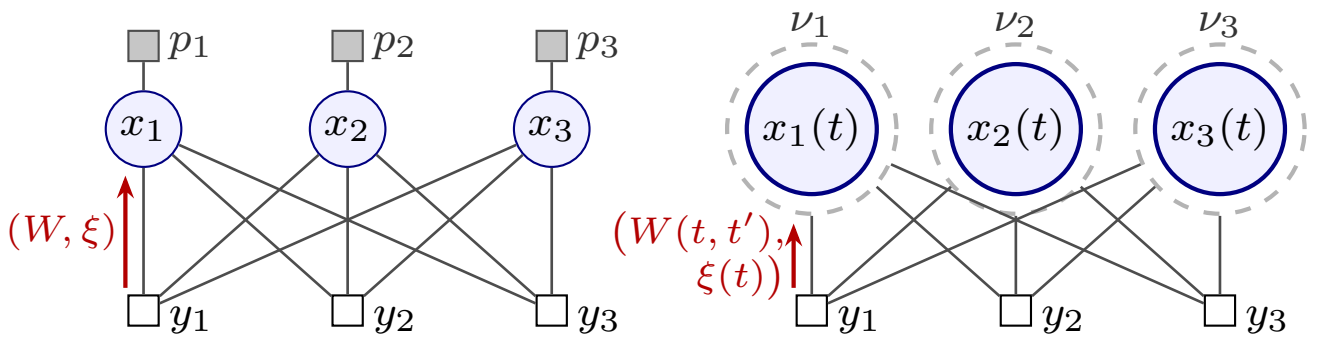
\begin{figure}[H]
\centering
\tikzset{
  fvar/.style={circle,draw=blue!50!black,thick,fill=blue!6,minimum size=8.6mm,inner sep=0},
  ref/.style={circle,draw=gray!60,dashed,thick,minimum size=11.2mm,inner sep=0},
  mfac/.style={rectangle,draw=black,fill=white,minimum size=2mm,inner sep=0},
  pfac/.style={rectangle,draw=gray!60!black,fill=gray!45,minimum size=2mm,inner sep=0},
  svar/.style={circle,draw=blue!50!black,fill=blue!6,minimum size=5mm,inner sep=0},
  edge/.style={draw=black!70,line width=0.5pt},
  msg/.style={-{Stealth[length=1.5mm]},red!70!black,line width=0.8pt},
}
\begin{tikzpicture}[font=\footnotesize]
\foreach \i/\x in {1/0.55,2/1.9,3/3.25}{
  \node[svar] (a\i) at (\x,0) {$x_\i$};
  \node[pfac] (ap\i) at (\x,0.55) {}; \draw[edge] (ap\i) -- (a\i);
  \node[gray!40!black,anchor=west,inner sep=1.5pt] at (ap\i.east) {$p_\i$};
  \node[mfac] (ay\i) at (\x,-1.2) {};
  \node[anchor=west,inner sep=1pt] at ($(ay\i.east)+(0.02,-0.06)$) {$y_\i$};
}
\foreach \i in {1,2,3}{\foreach \s in {1,2,3}{\draw[edge] (a\i) -- (ay\s);}}
\draw[msg] ($(ay1)+(-0.12,0.15)$) -- ($(a1.south)+(-0.12,-0.05)$)
   node[midway,left,red!70!black,inner sep=1pt,font=\scriptsize] {$(W,\xi)$};
\node[anchor=north] at (1.9,-1.33) {(a) $y_s=\sum_{i}a_{s,i}x_i+w_s$};
\begin{scope}[shift={(4.42,0)}]
\foreach \i/\x in {1/0.55,2/1.9,3/3.25}{
  \node[ref] (r\i) at (\x,0) {};
  \node[fvar] (b\i) at (\x,0) {$x_\i(t)$};
  \node[mfac] (by\i) at (\x,-1.2) {};
  \node[anchor=west,inner sep=1pt] at ($(by\i.east)+(0.02,-0.06)$) {$y_\i$};
  \node[gray!40!black,anchor=south,inner sep=0.8pt] at (r\i.north) {$\nu_\i$};
}
\foreach \i in {1,2,3}{\foreach \s in {1,2,3}{\draw[edge] (r\i) -- (by\s);}}
\draw[msg] ($(by1)+(-0.12,0.15)$) -- ($(r1.south)+(-0.12,-0.05)$)
   node[midway,left,red!70!black,inner sep=1pt,font=\scriptsize,align=right] {$\big(W(t,t'),$\\$\xi(t)\big)$};
\node[anchor=north] at (1.9,-1.33) {(b) $y_s=\sum_{i}\int_\T a_{s,i}x_i\,\dd t+w_s$};
\end{scope}
\end{tikzpicture}
\vspace{-2ex}
\caption{Factor graphs of a linear model for scalars (a) and functions (b), with priors $p_i$ (gray) and likelihoods. A function has no \ac{PDF}, so its prior $\nu_i$ is its reference (dashed), and messages $(W,\xi)$ become kernels and functions.}
\vspace{-2ex}
\label{fig:fg}
\end{figure}

Here, $\summ{x}$ is the ``not-sum''~\cite{KschischangFreyLoeliger2001} over all arguments of $f$ but $x$.
Every message thus excludes its destination, i.e., is extrinsic; on a tree, the product of all messages into $x_i$ is its unnormalized marginal.

The not-sum is a plain sum for discrete variables and an integral with $\dd x$ for real-valued ones, i.e., it weights all values equally.
We call this equal weighting the \emph{flat reference} of $x$: the weight $1$ per symbol of a finite alphabet, and the length $\dd x$ per small interval of $\R$, or the volume for vectors.
Every \acf{PDF} $p(x)$ is defined relative to it, since $\Pr\{x\in\mathcal A\}=\int_{\mathcal A}p(x)\,\dd x$, i.e., $p(x)$ is probability per unit length, just as a probability mass function is probability per symbol.
Note that the flat reference is not a prior, but only fixes how densities are measured, whereas the prior of $x$ is a separate factor, e.g., $p_i(x_i)$ in Fig.~\ref{fig:fg}(a).

\section{Functional Factor Graphs}\label{sec:ffg}
\subsection{From Vectors to Functions}\label{sec:vec}
Let each variable be a real function $x_i(t)$ on $\T$.
Computations then mirror those with vectors, with sums replaced by integrals, e.g., $Ax\triangleq\int_\T A(t,t')x(t')\,\dd t'$ in place of $\bA\bx$, compositions $\int_\T A(t,\tau)B(\tau,t')\,\dd\tau$ in place of products, the energy $\|x\|^2=\int_\T x(t)^2\,\dd t$, and the identity kernel $\delta(t-t')$ in place of $\mathbf I$, while $A^{-1}$ and $A^{1/2}$, if they exist, are the kernels whose compositions $A^{-1}A$ and $A^{1/2}A^{1/2}$ are $\delta$ and $A$.
The \emph{gain} $\sigma_1(A)\triangleq\sup_{u\neq0}\|Au\|/\|u\|$ and the \emph{energy} $\iintT A(t,t')^2\,\dd t\,\dd t'$ are the counterparts of the spectral and squared Frobenius norms, and a kernel has \emph{finite trace} $\int_\T A(t,t)\,\dd t$ if its singular values are summable.
Thus, gain $\le\sqrt{\text{energy}}\le$ sum of singular values, while $\delta$ has gain $1$, but infinite energy and trace.

\subsection{From PDFs to Likelihood Ratios}\label{sec:pdf}
To see why the \ac{SPA} fails for functions, let the prior of a function $x(t)$ be a zero-mean Gaussian process with a continuous covariance kernel $C(t,t')\triangleq\E\{x(t)x(t')\}$, whose orthonormal eigenfunctions $e_k(t)$, with eigenvalues $\lambda_1\ge\lambda_2\ge\dots>0$, i.e., $\int_\T C(t,t')e_k(t')\,\dd t'=\lambda_ke_k(t)$, span $L^2(\T)$, so $C(t,t')=\sum_{k=1}^{\infty}\lambda_ke_k(t)e_k(t')$, as for matrices.
Moreover, $\lambda_k\to0$, as the eigenvalues sum to the trace $\int_\T C(t,t)\,\dd t=\E\{\|x\|^2\}$, which is finite, as $C$ is continuous on the compact $\T$.
The coefficients $c_k\triangleq\int_\T x(t)e_k(t)\,\dd t$ are zero-mean and jointly Gaussian, with $\E\{c_kc_{k'}\}=\iintT e_k(t)C(t,t')e_{k'}(t')\,\dd t\,\dd t'$, which is $\lambda_k$ if $k'=k$ and $0$ otherwise, and thus independent, $c_k\sim\N(0,\lambda_k)$, which yields the Karhunen-Lo\`eve expansion
\begin{equation}\label{eq:KLE}
x(t)=\textstyle\sum_{k=1}^{\infty}c_ke_k(t)=\sum_{k=1}^{\infty}\sqrt{\lambda_k}\,\zeta_ke_k(t),
\end{equation}
with the \ac{iid}\ \emph{whitened coefficients} $\zeta_k\triangleq c_k/\sqrt{\lambda_k}\sim\N(0,1)$.

Just as $\mathbf z=\mathbf C^{-1/2}\bx=\sum_{k=1}^{K}\zeta_k\mathbf e_k$ whitens a vector $\bx\sim\N(\mathbf 0,\mathbf C)$, with $\mathbf C=\sum_{k=1}^{K}\lambda_k\mathbf e_k\mathbf e_k^\top$, by scaling its projections $c_k=\mathbf e_k^\top\bx$ to unit variance, the \emph{whitened process} of $x(t)$ is
\begin{equation}\label{eq:white}
z(t)\triangleq\int_\T C^{-1/2}(t,t')x(t')\,\dd t'=\textstyle\sum_{k=1}^{\infty}\zeta_ke_k(t),
\end{equation}
where $C^{-1/2}(t,t')\triangleq\sum_{k=1}^{\infty}\lambda_k^{-1/2}e_k(t)e_k(t')$ multiplies the $k$-th mode $c_ke_k(t)$ of $x(t)$ by $\lambda_k^{-1/2}$.
As the $\zeta_k$ are \ac{iid}\ with unit variance, $z(t)$ is unit white Gaussian noise, $\E\{z(t)z(t')\}=\sum_{k=1}^{\infty}e_k(t)e_k(t')=\delta(t-t')$, just as the outer products of an orthonormal basis of $\R^K$ sum to $\mathbf I$, but, as $\|z\|^2=\sum_{k=1}^{\infty}\zeta_k^2=\infty$, $z(t)$ is a formal object, and computations use only the $\zeta_k$, or whitened kernels such as $AC^{1/2}$.
Whitening thus turns every prior into the same unit white noise, whatever $C$, with precision $\delta$ in the role of $\mathbf I$.

Next, consider the \ac{PDF} of the $K$ strongest coefficients $c_1,\dots,c_K$, i.e., of the $K$-mode approximation of $x(t)$, which, as $c_k^2/\lambda_k=\zeta_k^2$, is
\begin{equation}\label{eq:pK}
p(c_1,\dots,c_K)=\textstyle\big(\prod_{k=1}^{K}(2\pi\lambda_k)^{-1/2}\big)\,e^{-\frac12\sum_{k=1}^{K}\zeta_k^2}.
\end{equation}

Its normalization grows faster than exponentially in $K$, as its logarithm $\frac12\sum_{k=1}^{K}\log(1/(2\pi\lambda_k))$ adds terms that grow without bound as $\lambda_k\to0$, whereas its exponential decays only as $\exp(-K/2)$, since $\frac1K\sum_{k=1}^{K}\zeta_k^2\to1$ by the law of large numbers.
At a function drawn from the prior, $p(c_1,\dots,c_K)\to\infty$ as $K\to\infty$ with probability one, i.e., the $K$-mode \acp{PDF}, which are probabilities per unit volume in $\R^K$, have no limit.
A function thus has no \ac{PDF}, as volume, and hence the flat reference, has no counterpart over infinitely many coefficients, i.e., no nonzero, locally finite, translation-invariant measure exists on $L^2(\T)$~\cite{Bogachev1998,DaPrato2006}, and the prior factors of the \ac{SPA} of Section~\ref{sec:kfl}, and their messages, have no limit either.

The likelihoods, however, do have a limit.
Let $\mathbf y\triangleq[y_1,\dots,y_S]^\top$ collect $S$ noisy observations of $x(t)$, with likelihood $p(\mathbf y|c_1,\dots,c_K)$, in which the other modes are averaged over their prior, and evidence $p(\mathbf y)$, in which all modes are.
By Bayes' rule, we have
\begin{equation}\label{eq:LR}
\frac{p(c_1,\dots,c_K|\mathbf y)}{p(c_1,\dots,c_K)}=\frac{p(\mathbf y|c_1,\dots,c_K)}{p(\mathbf y)},
\end{equation}
which does not contain the prior \ac{PDF}~\eqref{eq:pK}, whose diverging normalization and decaying exponential cancel, and, as a nonnegative martingale in $K$, converges with probability one to $p(\mathbf y|x)/p(\mathbf y)$, with $p(\mathbf y|x)$ the likelihood of the function.
We call this ratio of posterior to prior, or of likelihood to evidence, a \emph{likelihood ratio}, i.e., a density relative to the prior.

We therefore give each variable $x_i$ a \emph{reference measure} $\nu_i$, or \emph{reference}, relative to which its densities are defined: the flat reference, e.g., $\dd x_i$, for discrete, real- and vector-valued variables, whose priors $p_i(x_i)$ remain factors, and the prior for functions, whose densities become likelihood ratios.
In other words, a density is probability per unit of reference: per unit length, or per symbol, for flat references, and posterior probability per unit of prior probability for the prior.
Integrating any $q$ against a reference, $\int q\,\dd\nu_i$, means $\int q(x_i)\,\dd x_i$ for a flat reference, and the prior average $\E\{q(x_i)\}$ for a function, i.e., $\int q(x_i)p_i(x_i)\,\dd x_i$ if it had a \ac{PDF} $p_i$.

\begin{definition}[Functional factor graph]\label{def:ffg}
Let each variable $x_i$ take values in a set $\X_i$, such as a finite alphabet, $\R^d$ or a space of functions on $\T$, with a reference $\nu_i$.
A \emph{functional factor graph} is the factor graph~\cite{KschischangFreyLoeliger2001} of equation~\eqref{eq:g}, whose local functions $f_m$ may take functions as arguments.
Let $\nu$ be the product of all references.
If $g\ge0$ and $0<Z\triangleq\int g\,\dd\nu<\infty$, the graph represents the posterior
\begin{equation}\label{eq:P}
\Pr\{(x_1,\dots,x_I)\in\mathcal A\}=\tfrac1Z\textstyle\int_{\mathcal A}g\,\dd\nu,
\end{equation}
for every measurable $\mathcal A$, i.e., $g/Z$ is its density \emph{relative to} $\nu$.
\end{definition}

With flat references, $g/Z$ is the posterior \ac{PDF}~\cite{KschischangFreyLoeliger2001}, while, with the priors as references, they are no longer factors, $g$ collects the likelihoods, $Z$ is the evidence, i.e., the prior average of $g$, and equation~\eqref{eq:P} is Bayes' rule, with $g/Z$ the likelihood ratio~\eqref{eq:LR}.
Fig.~\ref{fig:fg}(b) shows this for the linear observation model $y_s=\sum_{i=1}^{I}\int_\T a_{s,i}(t)x_i(t)\,\dd t+w_s$, $s=1,\dots,S$, with weight functions $a_{s,i}(t)$, e.g., deltas for point sampling, and noise $w_s$, where each function is one node, whereas sampling first yields $K$ scalar nodes $x_i(k\Delta)$, $k=1,\dots,K$, per function, coupled by prior factors, with $\Delta\triangleq1/K$ for $\T=[0,1]$.

\vspace{-1ex}
\subsection{The Functional Sum-Product Algorithm}
The \ac{FSPA} keeps the variable-node rule~\eqref{eq:v2f}, and replaces the not-sum in the factor-node rule~\eqref{eq:f2v} by an integral over the other arguments, weighted by their references,
\begin{equation}\label{eq:ff2v}
\mu_{f\to x}(x)=\int f(X)\textstyle\prod_{x'\in n(f)\setminus\{x\}}\mu_{x'\to f}(x')\,\nu_{x'}(\dd x'),
\end{equation}
where $\nu_{x'}(\dd x')$ stands for $\dd x'$ with a flat reference, and for $p(x')\,\dd x'$ with a prior of \ac{PDF} $p$ in finite dimensions.
With the priors as references, equation~\eqref{eq:ff2v} is an average over independent draws of the other arguments $x'$ from their priors, i.e., an extrinsic likelihood of $x$.
As in~\cite{KschischangFreyLoeliger2001}, the \ac{FSPA} is exact on trees with nonnegative local functions, where the product of all messages into $x_i$ equals the integral of $g$ against the references of all other variables, i.e., $Z$ times the marginal posterior density of $x_i$ relative to its reference.\footnote{The proof of~\cite[App.~A]{KschischangFreyLoeliger2001} carries over, as the distributive law becomes an exchange of integration order, permitted by Tonelli's theorem since the joint reference is a product of $\sigma$-finite references; cf.~\cite{Zachary1983}.}

\vspace{-1ex}
\subsection{Implications of Function-Valued Variables}\label{sec:what}
Densities relative to a reference are standard~\cite{WainwrightJordan2008}, but, in message passing, taking the prior as reference has three practical consequences for functions, namely:

\emph{1) No uninformative message:} The all-ones ``unit message'' of~\cite{KschischangFreyLoeliger2001}, a constant density relative to the reference, carried no information, but now means ``prior only'', which, in Gaussian models, has zero precision and potential.
No message removes the prior, and every function needs a proper one, as functions have no flat reference, hence no flat prior.

\emph{2) Deterministic nodes must be eliminated:} Equality, sum and matrix nodes~\cite{Forney2001,Loeliger2004,Loeliger2007} are Dirac deltas, e.g., $\delta(x-x')$, whose sifting property, $\int\delta(x-x')\mu(x')\,\dd x'=\mu(x)$, holds relative to the flat reference only, as, relative to a prior with \ac{PDF} $p$, it would yield $\mu(x)p(x)$, and a function has no \ac{PDF}.
Consequently, hard constraints, e.g., copies of a function or $x_j=Ax_i$, must be substituted into the factors.

\emph{3) Messages must be admissible:} A message enters equation~\eqref{eq:ff2v} as a likelihood ratio relative to the prior, so it must be finite with probability one under the prior, and positive with positive probability, for the integral to be finite and positive, which we call \emph{admissible}.
For functions, this is restrictive, as, by the Feldman-H\'ajek theorem~\cite{Feldman1958,Hajek1958,Bogachev1998}, two Gaussian processes either have positive and finite likelihood ratios relative to each other, or are told apart without error by a single realization, e.g., covariances $C$ and $1.01\,C$, whose ratio is revealed by the infinitely many \ac{iid}\ whitened coefficients of a realization.
After whitening, a Gaussian message $\exp(-\frac12\iintT zWz+\int_\T\xi z)$ is admissible if its exponent is finite with probability one under the white prior, e.g., if $\xi$ has finite energy and $W$ finite trace, as $\iintT zWz$ then has mean $\tr W$, just as $\E\{\mathbf z^\top\mathbf W\mathbf z\}=\tr\mathbf W$ for a white vector $\mathbf z$, or $W$ finite energy if $\iintT zWz$ is centered~\cite{Bogachev1998}.
If, instead, $W=\epsilon\delta$, $\epsilon>0$, then $\iintT zWz=\epsilon\|z\|^2$ is infinite, as $z(t)$ has infinite energy, and the message is $\exp(-\infty)=0$ with probability one.
A message may change infinitely many mode variances, but only by decaying amounts, and may not add a second delta to the prior precision $\delta$, scaling all of them alike.

\section{Gaussian Functional Factor Graphs}\label{sec:gauss}
\subsection{Functional \texorpdfstring{\acs{GaBP}}{GaBP}}\label{sec:fgabp}
Let all priors $\nu_i$ be zero-mean Gaussian processes with covariance kernels $C_i$, and whiten the variables as in equation~\eqref{eq:white}, $z_i\triangleq C_i^{-1/2}x_i$.
A Gaussian factor of equation~\eqref{eq:g}, e.g., a likelihood, then reads, up to a constant, as
\begin{equation}\label{eq:gfac}
-\log f(X)=\tfrac12\textstyle\sum_{i,j}\iintT z_iQ^f_{ij}z_j-\sum_{i}\int_\T b^f_iz_i,
\end{equation}
where $i$ and $j$ run over the arguments of $f$, and the kernels $Q^f_{ij}$, with $Q^f_{ji}(t,t')=Q^f_{ij}(t',t)$, and the functions $b^f_i$ are the positive semidefinite precision and the potential, i.e., precision-weighted mean, contributed by $f$.
We assume that every $Q^f_{ii}$ has finite trace, and every $b^f_i$ finite energy, so the positive semidefinite $Q^f\triangleq[Q^f_{ij}]$ has finite trace, as do its blocks $Q^f_{ij}$, and every factor is finite with probability one under the prior.

For instance, the $S$ observations of Fig.~\ref{fig:fg}(b), with \ac{iid}\ noise $w_s\sim\N(0,\eta^2)$, satisfy $\int_\T a_{s,i}x_i=\int_\T\tilde a_{s,i}z_i$, with the whitened weight function $\tilde a_{s,i}\triangleq C_i^{1/2}a_{s,i}$, so their likelihoods $\exp(-\frac{1}{2\eta^2}(y_s-\sum_{i=1}^{I}\int_\T\tilde a_{s,i}z_i)^2)$ form one factor $f$, with
\begin{equation}\label{eq:lik}
\begin{aligned}
Q^f_{ij}(t,t')&=\textstyle\frac{1}{\eta^2}\sum_{s=1}^{S}\tilde a_{s,i}(t)\tilde a_{s,j}(t'),\\
b^f_i(t)&=\textstyle\frac{1}{\eta^2}\sum_{s=1}^{S}y_s\tilde a_{s,i}(t),
\end{aligned}
\end{equation}
i.e., the precision and potential of \ac{LMMSE} estimation.
Here, $Q^f_{ij}$ has rank at most $S$, and $\tilde a_{s,i}$ has finite energy even for point samples $a_{s,i}(t)=\beta_{s,i}\delta(t-t_s)$, with channel coefficients $\beta_{s,i}$, as $\tilde a_{s,i}(t)=\beta_{s,i}C_i^{1/2}(t,t_s)$ has energy $\beta_{s,i}^2C_i(t_s,t_s)$.

Collecting all factors yields the blocks $J_{ii}\triangleq\delta+\sum_fQ^f_{ii}$, $J_{ij}\triangleq\sum_fQ^f_{ij}$, $i\neq j$, and $h_i\triangleq\sum_fb^f_i$ of the posterior precision $J\triangleq[J_{ij}]$ and potential.
The prior precision $\delta$ is not a factor, but enters through the averaging in equation~\eqref{eq:ff2v}.
Regrouping $g$ into one unary factor per variable, with precision $J_{ii}-\delta$ and potential $h_i$, and one pairwise factor per pair $\{i,j\}$ with $J_{ij}\neq0$ yields a pairwise graph, loopy if the couplings form a cycle, on which the \ac{FSPA} computes Schur complements, i.e., it is \ac{GaBP}~\cite{WeissFreeman2001,Malioutov2006} with kernels in place of matrices, given by the functional updates
\begin{align}
W_{i\to j}&=-J_{ji}J^{-1}_{i\setminus j}J_{ij},\label{eq:fW}\\
\xi_{i\to j}&=-J_{ji}J^{-1}_{i\setminus j}\big(h_i+\textstyle\sum_{i'\in\partial i\setminus\{j\}}\xi_{i'\to i}\big),\label{eq:fxi}
\end{align}
where $J_{i\setminus j}\triangleq J_{ii}+\sum_{i'\in\partial i\setminus\{j\}}W_{i'\to i}$ is the precision collected by $z_i$ from all neighbors but $z_j$; $\partial i\triangleq\{i'\neq i:J_{ii'}\neq0\}$ are the neighbors of $z_i$; and $W_{i\to j}$ and $\xi_{i\to j}$ are the precision and potential of the message from $z_i$ to $z_j$.

We call equations~\eqref{eq:fW} and~\eqref{eq:fxi} \emph{functional \ac{GaBP}}.
Its belief of $z_i$ is Gaussian with mean $\hat z_i\triangleq\hat J_i^{-1}\hat h_i$ and covariance $\hat J_i^{-1}$, where $\hat J_i\triangleq J_{ii}+\sum_{i'\in\partial i}W_{i'\to i}$ and $\hat h_i\triangleq h_i+\sum_{i'\in\partial i}\xi_{i'\to i}$ collect the prior, the factors and all incoming messages, so the belief of $x_i$ has mean $C_i^{1/2}\hat z_i$ and covariance $C_i^{1/2}\hat J_i^{-1}C_i^{1/2}$, the latter approximate on loopy graphs, as in scalar \ac{GaBP}.
In practice, each $z_i$ is stored by $K$ coefficients in an orthonormal basis, and functional \ac{GaBP} runs as block \ac{GaBP} with $K\times K$ blocks, from zero messages, i.e., ``prior only''.
The change of reference thus amounts to whitening, but $J$ and the messages now have limits as $K\to\infty$.

\vspace{-1ex}
\subsection{Grid-Independent Convergence and Admissibility}\label{sec:conv}
\vspace{-1ex}

Scalar \ac{GaBP} converges if the model is \emph{walk-summable}~\cite{Malioutov2006}, i.e., if $\rho(|\mathbf R|)<1$, where $|\mathbf R|$ collects the absolute normalized couplings $|J_{ij}|/\sqrt{J_{ii}J_{jj}}$, $i\neq j$, with zeros on its diagonal.
For functions, let $R_{ij}\triangleq J_{ii}^{-1/2}J_{ij}J_{jj}^{-1/2}$, $i\neq j$, be the normalized couplings, and $\mathbf N$ the matrix with entries $N_{ij}\triangleq\sigma_1(R_{ij})$, the gain of the strongest mode by which $x_j$ acts on $x_i$, and $N_{ii}\triangleq0$, so $\mathbf N$ stays $I\times I$ for every discretization.
We call the model \emph{walk-summable} if $\rho(\mathbf N)<1$, as for scalars, where $\mathbf N=|\mathbf R|$.

\begin{theorem}[Grid-independent convergence]\label{thm:ws}
Let the precision of every factor be positive semidefinite, every $J_{ij}$ have finite gain, and every $h_i$ finite energy. If $\rho(\mathbf N)<1$, then \textup{1)} functional \ac{GaBP} started from zero messages converges, and its means $\hat z_i$ converge, with an error of $O(r^\ell)$ after $\ell$ iterations, for every $r>\rho(\mathbf N)$, to the solution $z_i^\star$ of $\sum_{j=1}^{I}J_{ij}z_j^\star=h_i$, $i=1,\dots,I$, which, under the assumptions of Section~\ref{sec:fgabp}, is the exact posterior mean; and \textup{2)} if each $z_i$ is stored by $K$ coefficients in an orthonormal basis, the discretized model has $\mathbf N_K\le\mathbf N$ entrywise, so $\rho(\mathbf N_K)\le\rho(\mathbf N)$, and block \ac{GaBP} with one block per function converges as in 1), for every $K$.
\end{theorem}

Theorem~\ref{thm:ws} follows from the walk-sum analysis of~\cite{Malioutov2006}, in its block form~\cite[Prop.~4.2.1]{MalioutovThesis2008}: after normalizing each $J_{ii}$ to $\delta$, i.e., replacing $z_i$ by $J_{ii}^{1/2}z_i$, the posterior precision is $\delta+R$, with $R$ collecting the $R_{ij}$, and its inverse, $\delta-R+R^2-\cdots$, sums over all walks $i_0\to i_1\to\cdots$ on the graph, each contributing a kernel of gain at most $N_{i_0i_1}N_{i_1i_2}\cdots$, so all walk-sums converge if $\rho(\mathbf N)<1$.
As the $\ell$-th iterate is exact on the depth-$\ell$ computation tree~\cite{WeissFreeman2001}, which contains all walks of length at most $\ell$ and whose coupling operator has gain at most $\rho(\mathbf N)$, every $J_{i\setminus j}$ has a bounded inverse, and the error is $O(r^\ell)$.
For 2), $N_{ij}=\sup_{u,v\neq0}|\iintT uJ_{ij}v|/(\|J_{ii}^{1/2}u\|\,\|J_{jj}^{1/2}v\|)$ is the strengthened Cauchy-Schwarz constant~\cite{Xu1992,GriebelOswald1995}, which can only decrease if $u$ and $v$ are restricted to $K$-dimensional subspaces, as in the mesh independence of Newton's method~\cite{Allgower1986}, so $K$ is chosen for accuracy only.
Block \ac{GaBP} is not new~\cite{MalioutovThesis2008,Kamper2019,Du2018,Fanaskov2022}, and, for vectors, $\rho(\mathbf N)<1$ is the block H-matrix condition~\cite{FeingoldVarga1962}; as gains are unchanged by invertible changes of variables within a block, Theorem~\ref{thm:ws} even holds for grid samples with exact priors, grouped by function, i.e., factoring first makes the condition grid-independent.

Under the assumptions of Section~\ref{sec:fgabp}, walk-summability also keeps all messages admissible, as every $W_{i\to j}$ in equation~\eqref{eq:fW} is a product of two finite-energy kernels and the finite-gain $J_{i\setminus j}^{-1}$, and thus has finite trace~\cite{ReedSimon1980}, while every $\xi_{i\to j}$ has finite energy.
Admissibility is, however, a separate requirement, as $\rho(\mathbf N)$ involves only gains, and a second delta (Section~\ref{sec:what}) has finite gain.

\vspace{-1ex}
\subsection{Grid \texorpdfstring{\acs{GaBP}}{GaBP} Under Refinement}\label{sec:grid}

\emph{Grid \ac{GaBP}} is scalar \ac{GaBP} from zero messages with one node per sample $x_i(k\Delta)$ at $K$ uniform grid points, where prior factors couple samples at most a fixed $\kappa$ grid steps apart, and finitely many point observations, each at one grid instant, couple the functions.
A \emph{round} is a set of message updates that use only messages of earlier rounds, the zero messages forming round~$0$, so any schedule is a sequence of rounds.

\begin{theorem}[Grid \ac{GaBP} under refinement]\label{thm:grid}
\textup{1)} After $\ell$ rounds, the estimate of $x_i(k\Delta)$ depends only on the observations within $\ell\kappa\Delta$ of $k\Delta$, so, for data with the covariance of the functional model, its mean-square deviation from the posterior mean $x^\star_i(k\Delta)$ is at least $\phi_i(k\Delta,\ell\kappa\Delta)$, where $\phi_i(t,\theta)$, the error of the best linear estimate of $x^\star_i(t)$ from the observations within $\theta$ of $t$, does not depend on $K$, is nonincreasing in $\theta$, and is positive if $x^\star_i(t)$ depends on a farther one, so an error below $\phi_i(t,\theta)$ takes more than $\theta K/\kappa$ rounds; \textup{2)} for a stationary first-order Gauss-Markov prior, $\kappa=1$ and $\rho(|\mathbf R|)\ge1-O(1/K^2)$; and \textup{3)} for a second-order prior whose interior grid precision rows are $\propto[1,-4,6,-4,1]$ up to $O(1/K)$, e.g., a double-pole autoregressive process of order two or a second-order random walk~\cite{RueHeld2005}, $\kappa=2$ and $\rho(|\mathbf R|)\ge\frac53-O(1/K)$.
\end{theorem}

\begin{figure*}[t]
\centering
\includegraphics[width=\textwidth]{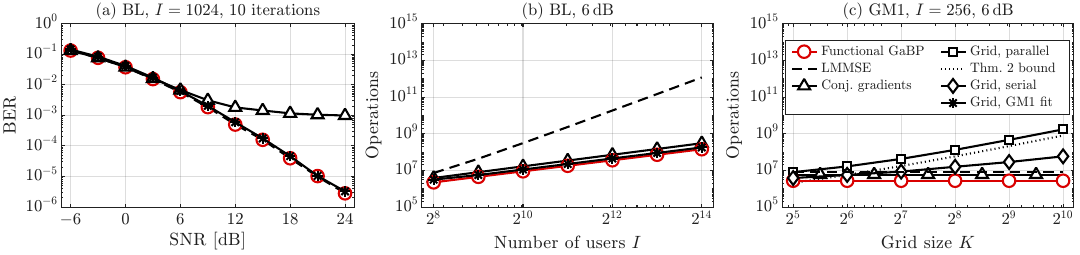}
\vspace{-4.5ex}
\caption{Multi-user detection: (a) bit error rate (BER) for \ac{BL}; arithmetic operations to convergence versus (b) $I$, for \ac{BL}, and (c) $K$, for GM1. Grid \ac{GaBP} uses the exact prior and the parallel schedule, unless marked. With the exact prior, grid \ac{GaBP} diverges for \ac{BL}, so (a) and (b) omit it.}
\vspace{-2ex}
\label{fig:num}
\end{figure*}

\emph{Proof sketch:} For 1), by induction on $\ell$, a message, or belief, of round $\ell$ depends on the data only through the potentials within $\ell$ edges, i.e., as prior edges span at most $\kappa$ grid steps and observation edges none, on the observations within $\ell\kappa\Delta$; the belief means are linear in these, so their error, set by the covariance alone, is at least that of the best linear estimate from them, which does not involve the grid.
For 2) and 3), the interior rows of the grid precision tend to $\propto[-1,2,-1]$ and $\propto[1,-4,6,-4,1]$, so the couplings of each sample to its nearest ones tend to $\frac12,\frac12$ and $\frac46,\frac46,\frac16,\frac16$, and all but $O(1)$ rows of $|\mathbf R|$ sum to almost $1$ and $\frac53$.
As $|\mathbf R|$ is symmetric and nonnegative, $\rho(|\mathbf R|)\ge\mathbf v^\top|\mathbf R|\mathbf v/\mathbf v^\top\mathbf v$, which is $\frac53-O(1/K)$ for the all-ones $\mathbf v$, cf.\ the thin-plate model of~\cite[Ex.~1]{MalioutovThesis2008}, and, for first-order priors and $v_k=([\mathbf J_i]_{kk})^{1/2}$ on function $i$ and zero elsewhere, whose block $\mathbf J_i$ of the grid precision has off-diagonal entries $\le0$, at least $1-\mathbf 1^\top\mathbf J_i\mathbf 1/\tr\mathbf J_i=1-O(1/K^2)$, as $\mathbf 1^\top\mathbf J_i\mathbf 1$ stays bounded, while $\tr\mathbf J_i$ grows as $K^2$.

\emph{Remark:} By 1), any schedule, or linear algorithm passing messages on the grid graph, e.g., Gauss-Seidel, needs $\Omega(K)$ rounds, so few serial sweeps suffice only if each takes $\Omega(K)$ rounds; by 2) and 3), refinement weakens walk-summability, which, being only sufficient, cannot prove divergence.

\vspace{-1ex}
\section{Example: Multi-User Detection of Waveforms}\label{sec:num}
\vspace{-1ex}
To evaluate the proposed scheme, consider an uplink, as in Fig.~\ref{fig:fg}(b), in which $I$ users spread their waveforms over $2I$ resources, e.g., subcarriers, each user occupying $4$ and each resource carrying $2$ users, at random, i.e., low-density spreading~\cite{GuoWang2008}, where, as in low-density parity-check codes, the graph is sparse, not the waveforms.
User $i$ transmits $x_i(t)=\sum_{k=1}^{2048}\sqrt{\lambda_k}\,\zeta_{i,k}\,e_k(t)$, $t\in[0,1]$, whose whitened coefficients~\eqref{eq:KLE}, for the covariance $C(t,t')$ below, are \ac{iid}\ equiprobable binary phase-shift keying symbols $\zeta_{i,k}\in\{\pm1\}$.
This eigenmode signaling, a stand-in for any waveform with this covariance, has, up to truncation, the prior covariance $C(t,t')=\sin(16\pi(t-t'))/(16\pi(t-t'))$ (bandlimited, \ac{BL}, with about $16$ significant eigenvalues), or $\exp(-\alpha|t-t'|)$ (first-order Gauss-Markov, GM1), with $\alpha=4$ in Fig.~\ref{fig:num}(c), for which grid \ac{GaBP} converges.

Each resource samples at the $32$ instants $\{1,\dots,32\}/32$, which yields $S=64I$ observations $y_s=\sum_{i=1}^{I}\beta_{s,i}x_i(t_s)+w_s$, i.e., $a_{s,i}(t)=\beta_{s,i}\delta(t-t_s)$, with $\beta_{s,i}\sim\N(0,1)$, \ac{iid}\ and constant over the frame if user $i$ occupies the resource of $y_s$, and $0$ otherwise, collected per resource in $\boldsymbol\beta_i\in\R^{2I}$; the \ac{SNR} is $1/\eta^2$, as $C(t,t)=1$.
Two functions are thus coupled only if they share a resource, so the functional factor graph has $I$ nodes with at most $4$ neighbors and is locally tree-like~\cite{MezardMontanari2009}, whereas sampling first yields $IK$ nodes.
The bits on the $8$ strongest modes, the others acting as interference, are detected as the signs of the posterior means of $\zeta_{i,k}$, computed, with the Gaussian prior of covariance $C$, by \ac{LMMSE}, which is exact; by functional \ac{GaBP}, which needs no truncation, as the likelihoods~\eqref{eq:lik} of a user lie in the span of the $32$ functions $C^{1/2}(t,t_s)$; and by grid \ac{GaBP} with $K=32$ for \ac{BL} and $K\in\{32,\dots,1024\}$, with the exact grid prior, for GM1, undamped, under a parallel or a serial forward-backward schedule.
As all users share the prior and the sampling instants, equation~\eqref{eq:lik} yields $J=\mathbf I_I\otimes\delta+(\mathbf H^\top\mathbf H/\eta^2)\otimes\Gamma$, with the sparse $\mathbf H\triangleq[\boldsymbol\beta_1,\dots,\boldsymbol\beta_I]$ and $\Gamma(t,t')\triangleq\sum_{s=1}^{32}C^{1/2}(t,t_s)C^{1/2}(t_s,t')$.
As $\Gamma$ has the eigenvalues $\gamma_1\ge\dots\ge\gamma_{32}$ of the covariance $\mathbf G\triangleq[C(t_s,t_{s'})]$ of $x(t)$ at these instants, functional \ac{GaBP} splits into $32$ scalar \ac{GaBP} detectors at the \acp{SNR} $\gamma_k/\eta^2$, and $\mathbf N$ is $|\mathbf R|$ of the first, at the \ac{SNR} $\gamma_1/\eta^2$, so one node per waveform adds no cost.

Fig.~\ref{fig:num}(b),~(c) count operations to convergence, i.e., until the squared deviation of the estimated $x_i(t_s)$ from their limit, pooled over $10$-$40$ trials, is below $10^{-4}$ of its energy; the limit is the \ac{LMMSE} estimate for every method but the GM1 fit.
A direct solve is costly, as random sparse graphs are expanders, so the Cholesky factors of $\mathbf I+\gamma_k\mathbf H^\top\mathbf H/\eta^2$ fill in, even in nested dissection.
For \ac{BL} at $6$\,dB and large $I$, the \ac{LMMSE} solve, split likewise, takes $0.27\,I^3$ operations, whereas functional \ac{GaBP}, with $\rho(\mathbf N)\le0.97$, needs $6$ iterations and $8.7\cdot10^3\,I$ operations, $8000$ times fewer at $I=2^{14}$, and conjugate gradients, $O(I)$ per iteration, but with global inner products, $19$-$25$ iterations and about twice the operations.

For \ac{BL}, sampling first fails with the exact prior, as $\mathbf G$ has eigenvalues from $2$ to below $10^{-15}$, so the samples, like a function, have, numerically, no \ac{PDF}, and grid \ac{GaBP}, which needs their precision, diverges for both schedules, with $\rho(|\mathbf R|)=5.4$ for $\mathbf G+10^{-6}\mathbf I$.
A GM1 grid prior fitted at adjacent instants, with $\alpha\approx14.5$, converges in $8$-$9$ iterations, with about $1.25$ times the operations of functional \ac{GaBP}, but solves another model, with $5$-$19\%$ more bit errors (Fig.~\ref{fig:num}(a), $I=1024$, $10$ iterations, $60$-$3000$ trials).
Functional \ac{GaBP}, which needs only $\mathbf G$, matches the \ac{LMMSE} \ac{BER}, within $0.7\%$ up to $18$\,dB and $3$ bit errors beyond, whereas conjugate gradients, in $10$ iterations, levels off at a \ac{BER} of about $10^{-3}$, $350$ times that of \ac{LMMSE} at $24$\,dB.

It can be seen from Fig.~\ref{fig:num}(c) that, for GM1 at $6$\,dB and ${I=256}$, grid \ac{GaBP} converges, unlike for \ac{BL}, but slows down: its $1-\rho(|\mathbf R|)$ falls as $1/K^2$, and, to converge, it needs over $0.4K$ parallel iterations ($2.2$ times, at $K=1024$, the bound $3K/16$ of Theorem~\ref{thm:grid}, as the summed $\phi_i(t_s,\theta)$ exceed the $10^{-4}$ threshold for ${\theta<3/16}$), or $11$-$14$ serial sweeps, whereas functional \ac{GaBP} needs $7$ iterations and fewer operations for every $K$, e.g., $600$ and $20$ times fewer, respectively, at $K=1024$, although $K=32$ is exact for GM1.

\vspace{-1.5ex}
\section{Conclusion}
\vspace{-1ex}
We offered a generalization of factor graphs and the \ac{SPA} to functions, with densities relative to the prior of each function, and answered the question of Section~\ref{sec:intro}: if $\rho(\mathbf N)<1$, \ac{GaBP} needs a bounded number of parallel iterations with one block per function, and grid \ac{GaBP} a growing one. Non-Gaussian factors, e.g., discrete symbols, are future work.

\bibliographystyle{IEEEtran}
\bibliography{references}
\end{document}